# The Bates Phase and Amplitude Monitoring System

D. Cheever, T. Ferrari, X. Geng, R. Page, T. Zwart
Bates Linear Accelerator Center, Middleton, MA 01949-1526

I. Introduction.

The MIT Bates LINAC is a 1 GeV electron accelerator used to perform precision measurements of the structure of nuclei and nucleons. The accelerator is a pulsed RF structure operating at 2856 MHz. Twelve CPI klystrons are situated along the length of the 200 m. LINAC. Each klystron is capable of delivering 6 MW of RF power for 30 us at a repetition rate of 600 Hz. Reliable operation of the machine requires that the relative phase of the 12 klystrons must be maintained to better than 1°. Similarly, the relative amplitude stability of each klystron must be maintained to better than $10^{-3}$.

To better meet these requirements we have constructed a RF phase and amplitude monitoring (PAM) system. This system samples a small fraction of the RF immediately downstream of each of the twelve klystrons. The complete system includes several main components: RF phase and amplitude detectors, analog signal conditioning, VME based digitizing and EPICS software for monitoring and calibration. The functionality and performance of each of these system components is described in detail below.

II. Hardware

Each of Bates' 6 transmitters house a pair of output klystrons and a drive klystron. These signals are available attenuated to under +30dBm via directional couplers on each klystron output waveguide and reference line taps. The RF signals are a patched to a chassis shown in figures 1 and 2 containing off-the-shelf 2.856 GHz RF components as listed in the appendix [1]. The output of these devices is impedance matched, fanned out, and routed to custom gated integrators and a custom video matrix switch. The integrators integrate and hold a gated portion of the pulsed signals and reset immediately before the next gate pulse, shown in Figure 3. The integrator outputs are low pass filtered at ~30Hz creating –10V to 10V ~DC signals that represents the peak amplitude or phase within the gate and are routed out of the transmitter gallery over multi-conductor twisted pair cabling to the 3 patch rooms where the non-synchronous VME ADC digitizes the signals. An Argonne Timing Board is used to set the relative timing of the reset and gate pulses, triggered by each transmitters RF "start" pulse. The VME DAC produces the ramping signals that shift the phase over 360º via the phase shifter to allow the necessary calibration. The pulsed phase and amplitude signals are also routed to the Bates video switch, enabling the control room to examine the signal for phase or amplitude ramps within a pulse. Shown on the following page as figure 4 is the front view of the RF detector chassis, integrators and signal fan-outs.

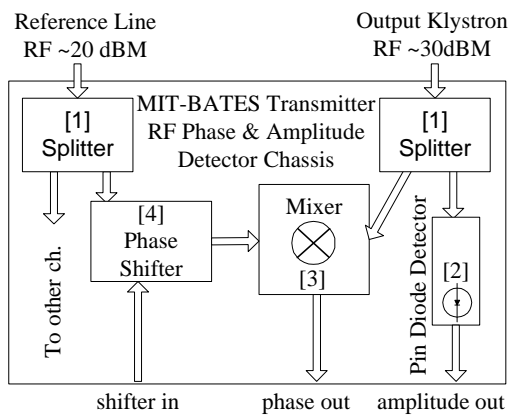

Figure 1. *RF Detection Hardware architecture*

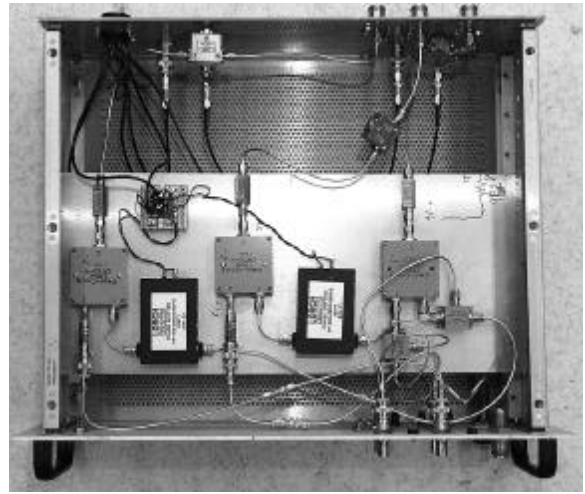

Figure 2. *RF chassis for two Klystrons*

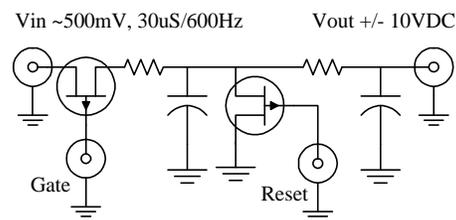

Figure 3. *Integrating Sample and Hold*

Specifications for the hardware are long term stability of displayed +/- 0.25° phase and +/- 0.1 Amps amplitude. Not shown is the VME crate that houses the ADC and timing board. These boards share a crate with other systems as is common. The integrator output is directly patched to the ADC via a custom BNC to ribbon cable interface board. The specifications were exceeded with the help of continuous calibration and special filtering as described in the software section.

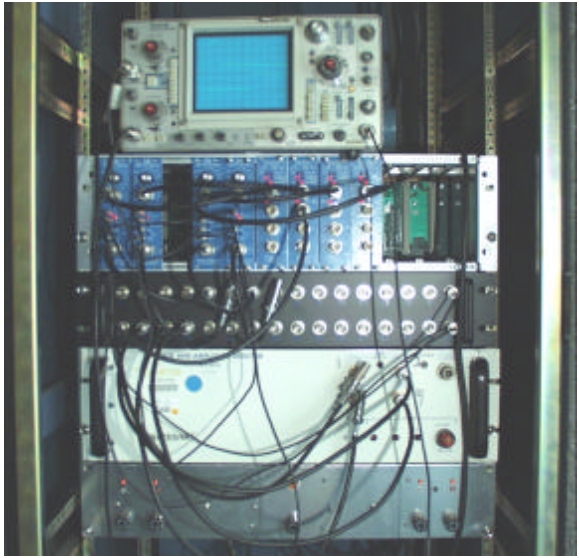

Figure 4. *Front view including integrators*

III. Software

The software of the PAM systems is comprised of a database and a sequencer program. The software functionality is described as follows.

1. Phase calibration.

Due to the use of passive mixers gain errors in the phase signal exist as the amplitude of the klystron signal is changed or drifts. To null these errors a full 360° phase sweep is performed at timed intervals. The sequencer program initiates a 0 to 10V DAC ramp that causes the phase shifter to swing the integrated phase signal thru a major part of the ADC input. Performing an arcSine with respect to peaks of the resulting sinusoid (the mixer peaks when differential phase of the two inputs are 0° and 180°) adequately nulls all common errors including RF cable temperature differences, chassis temperature drifts, and signal amplitudes. This calibration scheme eases the complexity and cost of temperature controlling the entire RF cable run and RF components. Calibration also extracts the slope of the phase shift, ensuring that drift are indicated in the right sign.

2. Zeroing

The end user of the PAM systems output is a machine operator. They are interested in drifts from a setting that earlier in the day resulted in good beam. Therefore they are interested in changes, not absolute data. The sequencer program allows for both phase and amplitude zeroing. The resultant variables are displayed as both strip charts and bar charts, with the former excellent at showing trends, and the former useful as an aid in returning the phase or amplitude of an individual Klystron to its previously known good "zeroed" value. These are shown below, and on the following page in Figure 7. Note the user has the ability to select specific klystrons for either phase or amplitude zeroing. Amplitude zeroing is simply the change from the last saved value, while phase zeroing causes the sequencer to proportionally change the phase shifter voltage until the ADC reads back the 0° point in the arcsine.

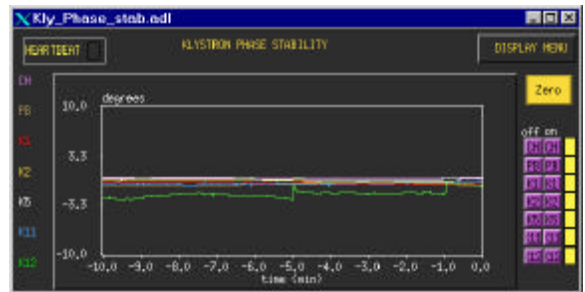

Figure 5. *PAM phase drift display*

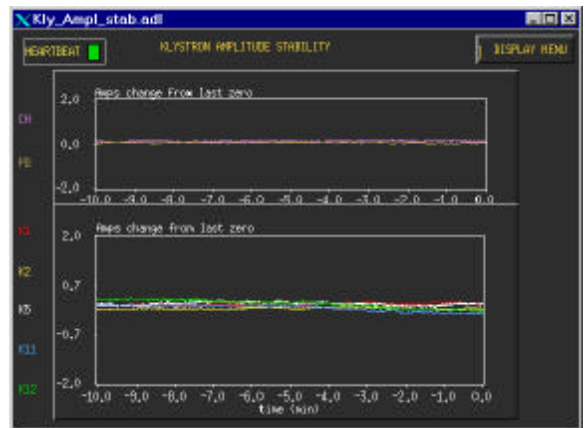

Figure 6. *PAM amplitude drift display*

3. Glitch handling.

Due to RF noise burst pickups a software glitch filter was necessary to meet these specifications as analog low pass filtering at the ADC and software smoothing proved unable to adequately eliminate the error contributed by the RF transients. The glitch filter is comprised of three additional records in the database, two "calc" and an "ao". The filter "histrograms" in the sense that it replaces the ADC raw value with the previous if it exceed a "glitch" threshold, but passes the value if it is under a "change" threshold. Software smoothing (moving average, as "SMOO" in an "AI" record) must be done in an additional subsequent "calc" record. The performance is shown below in Figure 8. The glitches are seen as >10 count events and are never greater than one acquisition long. The upper trace is the output on\f the glitch filter with a "glitch" setting of 5, and a "change" setting of 3. The smooth line is the output of the final "calc" record, with smoothing set to 9 old values and 1 new value (0.9).

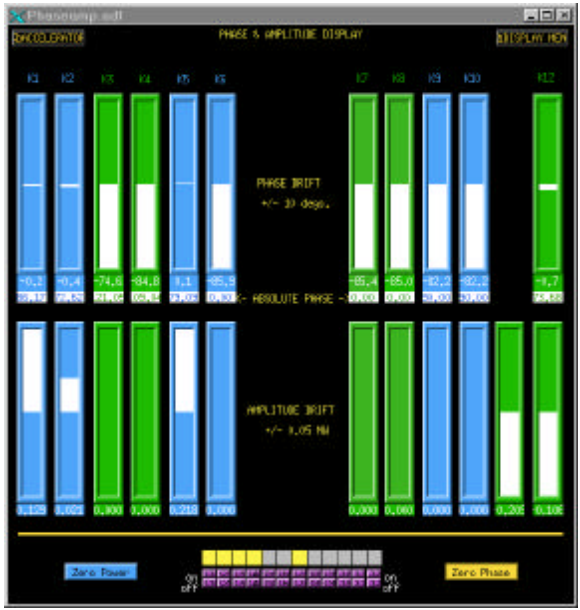

Figure 7. *PAM bar chart*

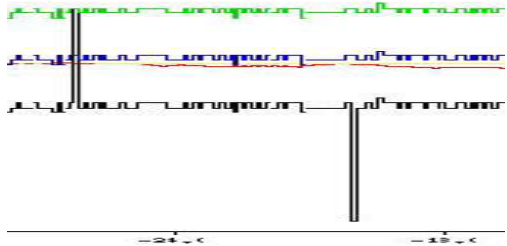

Figure 8. *Glitch handling.*

IV. Conclusions

The PAM system has been installed and operational over the last 3 ½ years in various forms. The current system has proven to have the resolution and stability to allow for the accurate monitoring of phase and amplitude drifts. Operationally the PAM system allows an operator to re-adjust either the amplitude or phase of the commonly lone drifting klystron rather than blindly re-tuning the beam via sequentially re-phasing all operating klystrons.

Over the past year PAM has proved enormously useful for catching step discontinuities in the unregulated high voltage applied to the klystrons, for monitoring and correcting phase drift of one or more klystrons and in establishing a clear correlation between a slow 3° phase drift on all klystrons and the performance of the primary water cooling system. This PAM information is presented to the operations group and accelerator scientists through the usual set of EPICS tools, MEDM strip charts, bar charts and archived data. This information has enabled the operations group to much more easily set and restore the phasing of the machine. It has allowed them to more effectively troubleshoot malfunctioning components of the RF system.

Future work at Bates for RF PAM systems will include implementing the EPICS alarm handler. The addition of 11 new digitally controlled phase shifters (one is already in use) will allow automatic phasing of the machine and potentially closed loop control of the RF phases for further suppression of drift and improved machine performance.

V. References

1. RF Detector Chassis Components

[1] Mini-circuits ZAPDQ-4         Splitter
[2] Hewlett Packard 8473D         Detector diode
[3] Mini-circuits ZEM-4300MH      Mixer
[4] Lorch VP-360-2800             Phase shifter
    Lorch 5BP7-2850/100-S         B.P. filter
    S.M.T. SM22052                Isolator